\documentclass[epj,referee]{svjour}
\usepackage{graphics}
\usepackage{bm}
\usepackage{graphicx,color}
\begin{document}
\title{A direct numerical simulation method for complex modulus of particle
dispersions}
\subtitle{}
\author{Takuya Iwashita\inst{1,2}\thanks{\emph{tiwashit@utk.edu}} \and  Takuya Kumagai\inst{1} \and Ryoichi Yamamoto\inst{1,2}\thanks{\emph{ryoichi@cheme.kyoto-u.ac.jp}}
}                     
\offprints{}          
\institute{Department of Chemical Engineering, Kyoto University - Kyoto 615-8510, Japan \and CREST, Japan Science and Technology Agency - Kawaguchi 332-0012, Japan}
\date{Received: date / Revised version: date}
%
\abstract{
We report an extension of the smoothed profile method (SPM)
 [Y. Nakayama, K. Kim, and R. Yamamoto, Eur. Phys. J. E {\bf 26}, 361
 (2008)], a direct numerical simulation method for calculating the
 complex modulus of the dispersion of particles, in which we introduce a
 temporally oscillatory external force into the system. The validity of
 the method was examined by evaluating the storage $G'(\omega)$ and loss
 $G''(\omega)$ moduli of a system composed of identical spherical
 particles dispersed in an incompressible Newtonian host fluid at volume
 fractions of $\Phi=0$, $0.41$, $0.46$, and $0.51$. The moduli were evaluated at
 several frequencies of shear flow; the shear flow used here has a
 zigzag profile, as is consistent with the usual periodic boundary
 conditions. The simulation results were compared with several experiments for colloidal dispersions of spherical particles.
} 
\maketitle
\section{Introduction}
\label{intro}
The viscoelastic properties of the dispersion of solid particles are of
particular importance in several scientific, engineering, and industrial
fields. These properties are strongly affected not only by direct
interactions between particles, but also by thermal fluctuations of the
system and the hydrodynamic interactions (HI) acting on dispersed
particles, mediated by the surrounding fluid \cite{ex1,ex2}. The
viscoelasticity of materials are most commonly characterized by the
complex modulus, which represents the mechanical response of materials
to temporally oscillating small shear deformations of frequency
$\omega$. The complex modulus consists of elastic and viscous components
({\it i.e.}, the storage modulus $G'(\omega)$ and the loss modulus
$G''(\omega)$). In general, materials tend to lose elastic energy by
dissipation when the frequency of external deformation is low, but can
store elastic energy when the frequency is high.

Understanding viscoelastic behavior has been the subject of both
fundamental and technological interest. 
Especially, the linear and non-linear responses of colloidal systems, including colloidal glasses and aggregated colloidal suspensions and gels,  
have attracted much attention in recent year \cite{Wyss,Vin,visex1}.
While extensive experimental
studies have been carried out to determine the viscoelastic properties
of particle dispersions over a wide range of volume fractions, from
dilute ($\Phi\simeq0$) to very dense ($\Phi\geq0.5$), theoretical
studies encounter a fundamental problem when the density of dispersion
is high, because of many-body effects and the long-range nature of the
HI among dispersed particles. Developing analytical theories becomes
even more difficult when the host fluids are complex, such as
electrolytes or viscoelastic media. Computer simulations can thus be
very powerful tools in theoretical investigations of dense dispersions
in general. 

The widely used simulation technique is Stokesian dynamics (SD) method \cite{SD},
which is based on the Stokes approximation (Re $\rightarrow$ 0), involves near-field lubrication forces and far-field many-body HIs. The SD method has succeeded in simulating the motions of particle dispersions at steady shear. However the viscoelasticity of particle dispersions at oscillatory shear has not been examined. Furthermore, the SD method is valid only for simulating the particles in a Newtonian fluid and is not applicable to the motions of particles in complex fluids, such as charged particles in electrolytes or viscoelastic media.

Various numerical methods have recently been proposed for simulating
dense dispersions, including some cases where the host fluids are
complex. In order to accurately track the motions of host fluids as well
as the motions of dispersed particles, a variety of numerical methods
have been developed. One of these methods, which we call the direct
numerical simulation (DNS) method for particle dispersions, involves
solving the Navier-Stokes (NS) equation for host fluids in a manner
consistent with boundary conditions defined according to the particle
positions and motions. Aside from several successful implementations of
DNS methods for particle dispersions 
\cite{hu,FPD,kajishima,naka1,naka2,iwa1,karn,IM0,IM1}, other methods
have been developed that do not rely on the Navier-Stokes equation to
resolve fluid motions. The coupling methods of particle dynamics with
the lattice-Boltzmann (LB) method \cite{LB} or stochastic rotational
dynamics (SRD) \cite{SR0,SR1} are the most popular alternatives to DNS,
because LB and SRD are believed to be more computationally efficient.

Problems studied here have very small Reynold number and  
the full NS equations for the host fluid are not necessarily solved to simulate the motions of particles in a Newtonian fluid. It is sufficient to use the methods based on the Stokes approximation. However there are several advantages for solving the NS equations for the host fluid. Those techniques mentioned above enable us to simulate
the short-time motions of particle dispersions, where the coupling between the particle motion and the fluid motion remains strongly, even if Reynolds number is very small. One of the typical behavior is the power-law decay in the velocity correlations of particles, which is known as "long-time tail''. In addition,  Those methods have succeeded in simulating several complex phenomena, such as the structure formation in colloidal gels where the pressure filed of the fluid plays an important role \cite{yama} and electrophoresis of charged colloidal dispersions \cite{kim1,araki}.

Although extensions of those methods have already been developed for
steady shear flow with DNS \cite{iwa4}, LB \cite{LB1,LB2,LB3},
 and SRD\cite{SR2}, there
exists no successful attempt for calculating the complex moduli of the
dispersions with an imposed oscillatory shear flow. 
The purpose of the present paper is to report
our successful reproductions of experimentally observed behaviors of the
storage $G'(\omega)$ and loss $G''(\omega)$ moduli in temporally
oscillating shear flow. 

To check the validity of the method, we first apply the method to a
Newtonian host fluid of viscosity $\eta$, which should exhibit a purely viscous response,
$G'(\omega)=0$ and $G''(\omega)=\eta\omega$. 
We then apply the method to dense dispersions composed of identical
spherical particles in a Newtonian host fluid. Since several
experimental measurements of the frequency dependence of $G'(\omega)$
and $G''(\omega)$ for dense colloidal dispersions have already been
reported \cite{visex1,visex2,mas,wata}, we compare our numerical results with those experiments, to
examine the validity of our method for dense particle dispersions.

\section{Simulation method}
\label{sec:1}
Let us consider a dispersion composed of identical spherical particles
of radius $a$ in a Newtonian host fluid subjected to oscillatory
shear. The host fluid is described by the velocity field $\bm v({\bm
r},t)$ and the pressure field $p(\bm r, t)$. The $\it i$th dispersed
particle is described by $\{\bm R_i, \bm V_i, \bm \Omega_i\}$, where
$\bm R_i$ is the position of the particle, $\bm V_i$ is the
translational velocity, and $\bm \Omega_i$ is the rotational
velocity. The coupling scheme between the fluid motions and the particle
motions is based on the smoothed profile method (SPM), which introduces
a particle density field $\phi(\bm r, t)\in [0:1]$ on the entire field,
$\phi=1$ for the particle domains and $\phi=0$ for the fluid
domains. These domains are separated by a thin interfacial domain of thickness $\xi$. The SPM is an efficient method to resolve the HIs between the
fluid and particle motions; its details are given in \cite{naka1,naka2,karn}.

The time evolution of the host fluid is governed by the Navier-Stokes
equation
%
\begin{equation}
\rho_f(\partial_t {\bm v} + {\bm v}\cdot \nabla {\bm v}) = \nabla \cdot\bm \sigma
 +\rho_f\phi{\bm f_p} +\rho_f \bm f^{shear}
\label{nseq}
\end{equation}
%
with the incompressibility condition $\nabla \cdot \bm v=0$, 
where $\rho_f$ is the density of the fluid, $\eta$ is the shear
viscosity, the stress tensor $\bm \sigma = -p\bm I + \eta\{\nabla \bm v
+ (\nabla \bm v)^T\}$, and $\bm f^{shear}(\bm r, t)$ is an external
force field that is introduced to enforce an oscillatory shear flow on
the entire system. The
body force $\phi \bm f_p$ is introdued to ensures the rigidity of particles and the appropriate boundary condition at the fluid/particle interface \cite{naka1,naka2,karn}.

The time evolution of the $\it i$th particle with mass $M_i$ and moment
of inertia $\bm I_i$ is governed by Newton's equations of motion:
\begin{eqnarray}
M_i \dot {\bm V_i} &=& {\bm F^H_i} + {\bm F^C_i} + {\bm G_i^V},\ \ \
\dot {\bm R_i} = {\bm V_i},\\
{\bm I_i}\cdot \dot{\bm \Omega_i} &=& {\bm N^H_i} + {\bm G_i^\Omega},
\end{eqnarray}
where $\bm F^H_i$ and $\bm N^H_i$ are the hydrodynamic forces and
torques, respectively, exerted by the host fluid on the particle. $\bm
F_i^C$ is the repulsive force that prevents the particles from
overlapping, and a truncated Lennard-Jones potential,
$V(r_{ij})=4[(\sigma/r_{ij})^{36}-(\sigma/r_{ij})^{18}+1/4]$ for
$r_{ij}<2^{1/18}\sigma$  or $V(r_{ij})=0$, is adopted in this
work. Here, $\sigma=2a$ and $r_{ij}=|\bm R_i - \bm R_j|$. $\bm G_i^V$
and $\bm G_i^\Omega$ are the random force and torque, respectively, due
to thermal fluctuations. These random fluctuations are assumed to be
$\langle \bm G_i^V\rangle=\langle \bm G_i^\Omega\rangle=0$, $\langle \bm
G_i^{V}(t)\cdot\bm G_i^{V}(0)\rangle=3\alpha_V\delta(t)$ and $\langle
\bm G_i^{\Omega}(t)\cdot\bm
G_i^{\Omega}(0)\rangle=3\alpha_\Omega\delta(t)$, where $\langle \rangle$
denotes time averaging, and $\alpha_V$ and $\alpha_\Omega$ are numerical
parameters to control the particle temperature $T$. The procedure for
determining the temperature is described in \cite{iwa3}.

The apparent stress $\bm \sigma^{app}$ of the dispersion is written as
\begin{equation}
%
\bm \sigma^{app} = \frac{1}{V}\int d\bm x \bm x \rho_t \bm f^{shear}
- \frac{1}{V}\int d\bm x \bm x\frac{d}{dt}(\rho_t \bm v)\label{sigma}
\end{equation}
where $\rho_t=(1-\phi)\rho_f + \phi\rho_p$, $\rho_p$ is the density of
the particles, and $V$ is the volume of the system. The derivation of
Eq. (\ref{sigma}) was reported in \cite{iwa4}.

The apparent stress $\bm \sigma^{app}$ consists of two terms: the first
term is a stress tensor including the external force and the second term
is a stress tensor arising from the acceleration of the dispersion. In
experimental viscoelastic measurements of dispersions, the acceleration
term can be ignored, because the relaxation time scales related to the
acceleration are considerably smaller than the experimental time
scales. On the other hand, in the simulations based on the DNS approach,
the acceleration term strongly affects the apparent stress of the
dispersion.

There are two key points for calculating the apparent stress $\bm
\sigma^{app}$ of the dispersion: i) how to calculate the acceleration
term in $\bm \sigma^{app}$, and ii) how to introduce the external force
$\bm f^{shear}$. The first point is straightforward. For a simulation time step $h$, the acceleration term
can be simply calculated as $1/V\int d\bm x \bm x(\rho_t\bm v(\bm x,
t+h)-\rho_t\bm v(\bm x, t))/h$.
Next the external force $\bm f^{shear}$ is introduced as a body force, to
enforce the following oscillatory velocity field over the entire system,

\begin{eqnarray}
v^0_{x}(y) &=& \left \{
\begin{array}{ll}
\dot\gamma (t) (-y - L_y/2), &(-L_y/2<y\leq -L_y/4) \\
\dot\gamma (t) y, &(-L_y/4<y\leq L_y/4) \\
\dot\gamma (t)(-y + L_y/2) &(L_y/4<y\leq L_y/2) \\
\end{array}
\right.
\\
\dot\gamma (t) &=& \dot\gamma_0\cos(\omega t)
\end{eqnarray} 
where $y$ denotes distance in the velocity-gradient direction, and
$\dot\gamma(t)$ is the  oscillatory shear rate, with amplitude
$\dot\gamma_0$ and frequency $\omega$. Here, the flow is imposed in the
$x$ direction and $L_y$ is the length of the system in the $y$
direction. This zigzag velocity profile was first used to simulate dispersions
in a steady shear \cite{iwa3,iwa4}. Note that $\bm f^{shear}$ is
defined to be a body force that constrains the velocity field of the
dispersions, and its explicit form is written as 
\begin{equation}
\bm f^{shear}=(\rho_t(v^0_{x}(\bm r) - v_x (\bm r))/h, 0, 0), \bm r\in V.
\end{equation}

If the external force is introduced as a boundary force, then the
development of the velocity from the boundary to bulk ({\it i.e.}, the
propagation modes) are observed for short time scales, such as the
kinematic time scale $\tau_\nu (=\rho_f a^2/\eta)$. In the simulation,
the propagation modes influence the overall viscoelastic properties of
the dispersion. On the other hand, in experimental measurements the
propagation modes are ignored. Using the external forces mentioned
above, we can eliminate the propagation modes numerically.

To measure the storage modulus $G'(\omega)$ and loss modulus
$G''(\omega)$, we monitor the $xy$ component of the apparent shear
stress, $\sigma^{app}_{xy}$, and shear rate $\dot\gamma$ as a function
of time. In general, the $xy$ component of the stress is written as 
\begin{equation}
\sigma^{app}_{xy}=\sigma_0\cos(\omega t - \delta)
\end{equation}
where $\sigma_0$ is the amplitude of the stress and $\delta$ is the
phase difference between $\sigma^{app}_{xy}$ and $\dot\gamma$. In our
model, the shear rate is an externally controlled parameter. By using
the obtained $\sigma_0$ and $\delta$, we can determine the following
dynamic viscoelastic moduli:
\begin{equation}
G'(\omega) = \frac{\sigma_0 \sin\delta}{\dot\gamma_0}\omega, \ 
G''(\omega)=\frac{\sigma_0 \cos\delta}{\dot\gamma_0}\omega.\label{moduli}
\end{equation} 

\section{Simulation results}
Three-dimensional simulations were performed at several frequencies ranging from 0.0005 to 0.2. 
The amplitude of strain,  $\gamma_{0}
=\int_0^{\pi/2\omega}\dot\gamma(s)ds=\dot\gamma_0/\omega$, was set to be
0.2. The spatial and temporal units are expressed in terms of the
lattice spacing $\Delta$ and  $\rho_f\Delta^2/\eta$, respectively. The
system lengths are $L_x=L_y=L_z=64$. The parameters of the simulation
are $a=4$, $\xi=2$, $\rho_p=1$, $\rho_f=1$, $\eta=1$, and $k_BT=7$. The
dispersed particles are initially randomly distributed. 

If we consider a particle of radius $0.4 \mu$m in water at room temperature, the units of space and time correspond to be $0.1 \mu$m and $0.0112\  \mu$sec, respectively. In this case the simulated range of the frequency, $f=\omega/2\pi$, is between 7.1 and 2842 kHz. 
\subsection{Test of the simulation method}
\begin{figure}
\includegraphics[scale=.8]{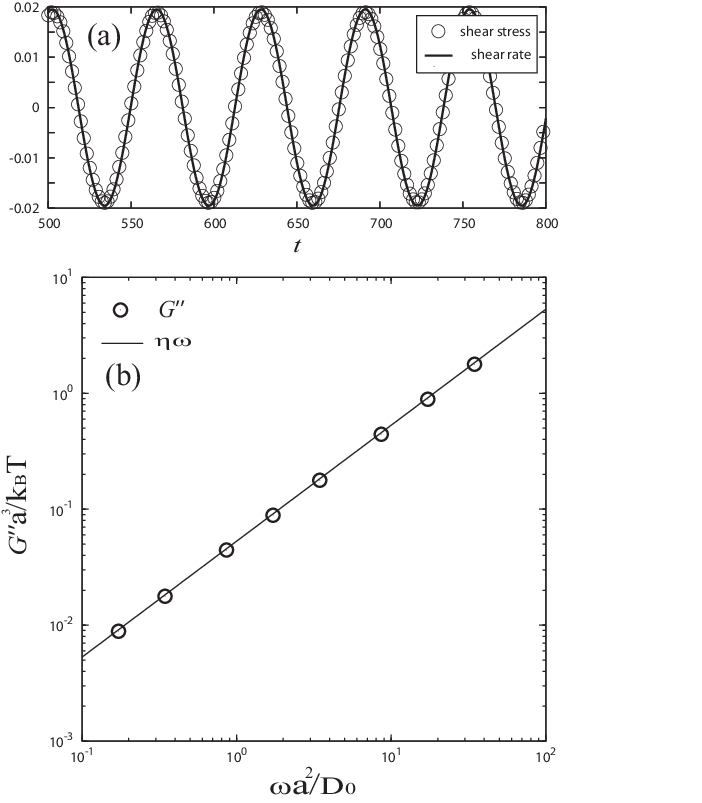}
\caption{\label{fig1}
The complex muduli of the host fluid at $\Phi=0$: (a)
 Time evolutions of the shear stress $\sigma^{app}_{xy}$ and the shear
 rate $\dot\gamma$ at $\omega=0.1$. (b) Frequency dependence of the loss
 modulus $G''$ of the host fluid. The modulus and frequency are scaled by $a^3/k_BT$ and $a^2/D_0$, respectively, and $D_0=k_BT/6\pi\eta a$. The solid line represents the loss
 modulus of the host fluid, $\eta\omega$.
}
\end{figure}

In order to test the validity of our method, we applied the method to a
Newtonian host fluid that does not contain dispersed particles. Figure
\ref{fig1} (a) shows the time evolutions of the shear rate $\dot\gamma$
and shear stress $\sigma^{app}_{xy}$ at $\omega=0.1$. It can be seen
that both curves develop in time with the same phase, $\it i.e.$,
$\delta=0$. This behavior represents the typical features of Newtonian
fluids. The loss modulus $G''$ was then calculated using
Eq. (\ref{moduli}) for different frequencies. Figure \ref{fig1} (b)
displays the frequency dependence of the loss modulus $G''$ for the host
fluid. The modulus and frequency are non-dimensionalized by $a^3/k_BT$ and $a^2/D_0$, respectively, where $D_0$ is the self-diffusion coefficient of a Brownian particle at infinite dilution, $D_0=k_BT/6\pi\eta a$. The loss modulus increases linearly with $\omega$, and its slope
is equal to the viscosity of the host fluid. The solid line indicates
the loss modulus of the host fluid, $\eta\omega$. These results show
that the viscoelastic properties of the host fluid are correctly
reproduced by the simulation.

\begin{figure}
\includegraphics[scale=.8]{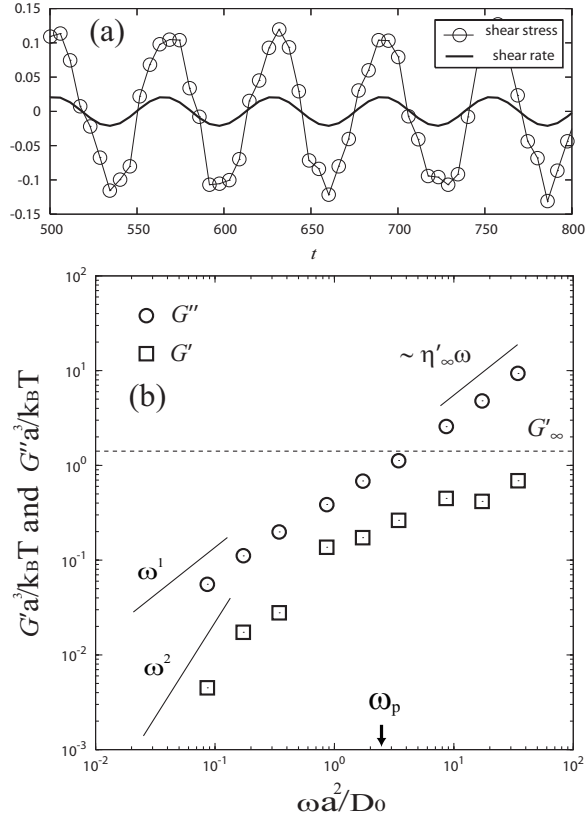}
\caption{\label{fig2}
The complex moduli of the dispersion at $\Phi=0.41$:
 (a) Time evolutions of the shear stress $\sigma^{app}_{xy}$ and the
 shear rate $\dot\gamma$ at $\omega=0.1$. (b) Frequency dependence of
 the storage modulus $G'$ ($\bigcirc$) and loss modulus $G''$
 ( ) for the dispersion. The modulus and frequency are scaled by $a^3/k_BT$ and $a^2/D_0$, respectively.
$G'_{\infty}$ is the high frequency elastic modulus, which was obtained by the numerical integration of a theoretical expression for $G'_{\infty}$ (dotted line) \cite{Russel}, and $\eta'_{\infty}$ is the high frequency viscosity. The down arrow indicates the inverse of the
 particle relaxation time, $\omega_p=\tau_p^{-1}=6D_0(\Phi)/a^2$.
}
\end{figure}

We next investigated a concentrated dispersion composed of spherical
particles fluctuating in the host fluid. The volume fraction of the
dispersed particles was $\Phi=0.41$. Figure \ref{fig2} (a) shows the
time evolutions of the shear rate $\dot\gamma$ and shear stress
$\sigma^{app}_{xy}$ at $\omega=0.1$ and $\Phi=0.41$. The behavior of the
shear stress of the dispersion differs from that of the host fluid shown
in Fig. \ref{fig1}. We can see that there are phase differences between
the shear rate and shear stress ({\it i.e.}, $\delta \neq 0$), and the
amplitude of the shear stress becomes greater than that of the shear
rate.

From the obtained $\delta$, $\sigma_0$ and $\dot\gamma_0$, we can
calculate the storage modulus $G'$ and the loss modulus $G''$ at
different frequencies. Figure \ref{fig2} (b) shows the frequency
dependence of the storage modulus and the loss modulus for the
dispersion at $\Phi=0.41$. The modulus and frequency are scaled by $a^3/k_BT$ and $a^2/D_0$, respectively. For low frequencies, $G'(\omega)$ increases
linearly with $\omega^2$ and $G''(\omega)$ increases linearly with
$\omega$. As $\omega$ increases, $G'$ grows monotonically until it
reaches a plateau region, while $G''$ develops up to a linear region,
and its slope is smaller than the slope at low frequencies. 
The slope at high frequencies represents the high frequency viscosity $\eta'_\infty$, which is given by
\begin{equation}
\eta'_\infty = \lim_{\omega \rightarrow \infty} G''(\omega)/\omega.
\end{equation}
From Fig. \ref{fig2} the high frequency viscosity was roughly estimated as $\eta'_\infty \simeq 5.2$.
Next we focus on the high-frequency elastic shear modulus $G'_\infty$, which is defined as
\begin{equation}
\lim_{\omega \rightarrow \infty} G'(\omega) = G'_{\infty}.
\end{equation}
We compared our results with a theoretical expression for $G'_{\infty}$ of hard spheres, which has been derived by Lionberger and Russel \cite{Russel}. To evaluate the theoretical value, we performed a numerical integration by using the obtained $\eta'_\infty$ and Percus-Yevick distribution function for $\Phi=0.41$. The calculated high-frequency modulus, $G'_\infty$, is shown in Fig. \ref{fig2}. We can see the values of $G'$ at high frequencies approach to the theoretical value. 

Furthermore, the loss modulus is larger than the storage modulus, and the dispersion
behaves like a viscous fluid. Here, we define the particle relaxation
time to be $\tau_p=a^2/6D_0(\Phi)$, where $D_0 (\Phi)$ is the diffusion
coefficient of a spherical particle in the dispersion at thermal
equilibrium. From the equilibrium calculations ($\dot\gamma=0$), we can
estimate the diffusion coefficient at $\Phi=0.41$, resulting in
$\omega_p\equiv \tau_p^{-1}\sim 1.5\times 10^{-2}$. The characteristic frequency $\omega_p$ is also indicated as an arrow in Fig. \ref{fig2}. At $\omega \sim \omega_p \ (\omega\tau_p\sim 1)$, the
onset of elasticity is clearly observed. These behaviors accurately
represent the typical viscoelastic features of concentrated dispersions
\cite{visex1}. 

\begin{figure}
\includegraphics[scale=.8]{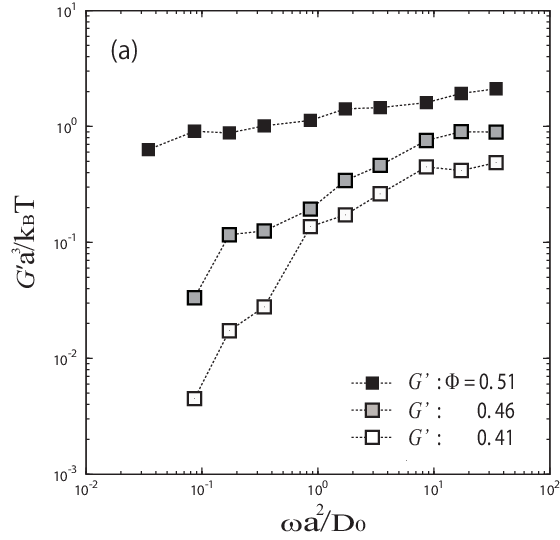}
\includegraphics[scale=.8]{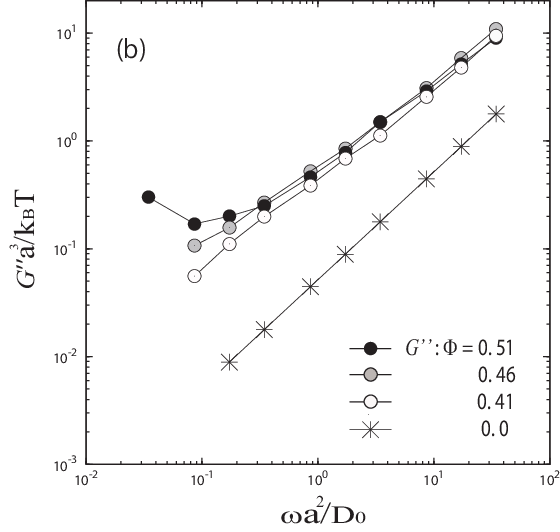}
\caption{\label{fig3}
Frequency dependence of (a) the storage modulus $G'$ and (b) the loss modulus $G''$
 at different frequencies: $\Phi=0.51$ (black symbols), $\Phi=0.46$ (gray symbols),  
$\Phi=0.41$ (white symbols), and $\Phi=0.0$ ($\ast $). The modulus and frequency are scaled by $a^3/k_BT$ and $a^2/D_0$, respectively. The amplitude of strain
 is set to 0.2.
}
\end{figure}

Finally, we examined the dispersion behavior of particles at higher
concentrations, $\Phi=0.46$ and $0.51$. The frequency dependence of $G'$ and $G''$
for different volume fractions is shown in Fig. \ref{fig3}; the data at $\Phi=0$ and $0.41$ are also plotted for comparison.
For $\Phi=0.46$, the structures of the particles are still randomly distributed under shear, wheres for $\Phi=0.51$ the system 
forms crystallize phases completely.
For $\Phi=0.46$ and $0.51$, each modulus shifts to higher values than those at $\Phi=0.41$. 
At low frequencies, the $\Phi$ dependence of both moduli is remarkable: the
storage modulus $G'$ rises rapidly with increasing $\Phi$, causing the
storage modulus to become larger than the loss modulus at low
frequencies and $\Phi=0.51$. 
The dispersion then becomes elastic. Furthermore, the loss
modulus $G''$ at $\Phi=0.51$ is a concave curve, with a minimum at low
frequencies. 

We found that the viscoelastic properties of the concentrated
dispersions depend strongly on the volume fraction. In addition, at
$\Phi=0.51$, the storage modulus increases very slowly with increasing
$\omega$, and at $\omega=0.02$, a crossover from elastic- to
viscous-dominant regions is observed. 

\subsection{Comparison of simulations and experiments}
\begin{figure}[h]
\begin{center}
\includegraphics[scale=.8]{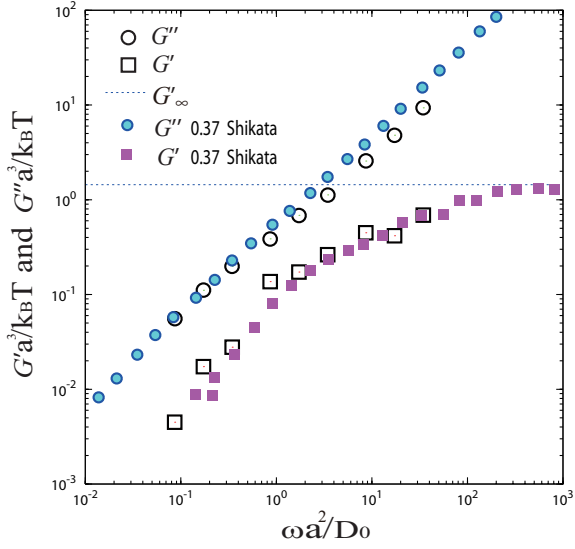}
\caption{\label{fig4} The frequency dependent shear moduli of the simulations at $\Phi=0.41$ (open symbol) and the experimental results at $\Phi=0.37$ for 53 nm silica particles in ethylene glycol/glycol (filled symbol), which were measured by Shikata and Pearson \cite{visex1}, and the high frequency elastic modulus from the simulation (dotted line).
}
\end{center}
\end{figure}

The simulation results obtained in this work were compared with experimental results for the dynamic viscoelasticity of colloidal dispersions, which were measured by several group \cite{visex1,visex2,mas,wata}. These experiments use colloidal dispersions of spherical particles dispersed in solvent. All experimental data were also plotted as a function of dimensionless variables, $G' a^3/k_BT$, $G'' a^3/k_BT$, and $\omega a^2/D_0$.  In Fig.\ref{fig4}, the simulation results for elastic and loss modulus at $\Phi=0.41$ were plotted together with those measured experimentally by Shikata and Pearson \cite{visex1}. No fitting parameters are used. The viscoelastic responses of the simulations agree well with those of the experiment,  although the volume fraction between them is not exactly the same. This clearly confirms our simulation can provide the typical viscoelastic behavior of colloidal dispersions in fluid states.

\begin{figure}[h]
\begin{center}
\includegraphics[scale=.8]{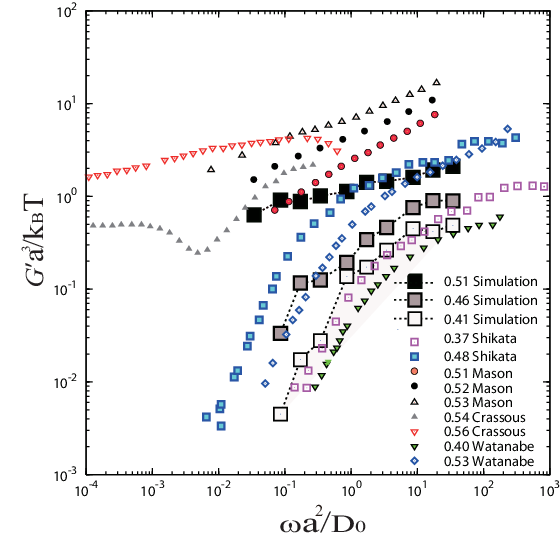}
\includegraphics[scale=.8]{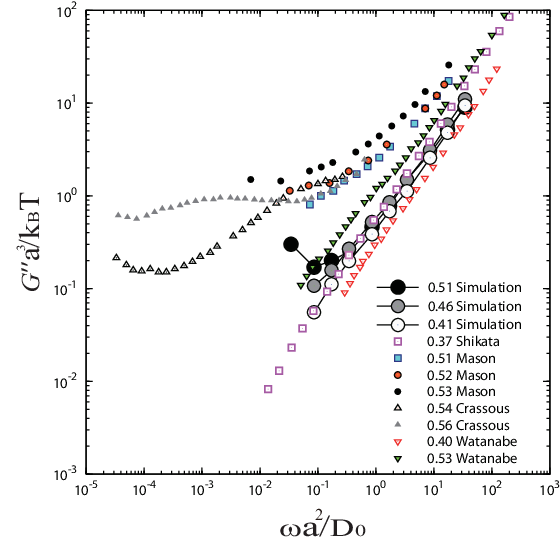}
\caption{\label{fig5} Data for the elastic and loss modulus of colloidal dispersions in an oscillatory shear flow. The experimental data were collected from several papers (Shikata and Pearson \cite{visex1}; 53 nm silica sphere in ethylene glycol/glycol, Mason \cite{mas}; 210 nm silica sphere in ethylene glycol, Crassous \cite{visex2}; thermosensitive particles in water, and Watanabe \cite{wata}; 46 nm silica sphere in ethylene glycol/glycol). The numbers indicate the volume fraction of particles.
}
\end{center}
\end{figure}

Comparison of simulation results and several experiments, including concentrated dispersions with higher $\Phi$, would give us a comprehensive information for the understanding of the dynamical behavior of colloidal dispersions. Figure \ref{fig5} show the experimental data on the elastic and loss modulus of concentrated dispersions over a wide range of volume fractions from 0.37 to 0.56 \cite{visex1,visex2,mas,wata}, and  our simulation results were also plotted.  These experimental data were measured over a range of $\omega a^2/D_0$ from $10^{-5}$ to $10^{3}$, and our simulation results lie within the range. The experimental curve shifts at  lower values of $\omega a^2/D_0$ as the volume fraction increases and this reflects the characteristic time scales of system become longer, and the dynamics of dispersed particles slows down.

The experimental data for high volume fraction vary widely depending on each experiment, and it is considered that the viscoelastic response becomes sensitive to the details of direct interactions between particles. The repulsive potential used in this study can produce the features of dispersions with relatively small volume fraction (See Fig. \ref{fig4}). The simulation at $\Phi=0.51$ has a crystalline structure over the whole frequency range, where a colloidal cystal is formed. Thus the simulation results at $\Phi=0.51$ are completely different from the experiments, for which the structures are in fluid and glasslike states. 
Recently Crassous {\it et al} \cite{visex2} have examined the effect of crystallization on the viscoelasticity of the dispersions at low frequency regime in the vicinity of the glass transition, and  they found the elastic modulus is greater than the loss modulus at low frequencies even in fluidlike states. This behavior is qualitatively similar to the simulation results at $\Phi=0.51$, and it represents solidlike respnses.. In the present work the dispersions in glassy states were not studied.
\section{Conclusion}
We have developed a DNS method for simulating dynamics of solid
particles dispersed in simple and complex fluids
\cite{naka1,naka2}. This method, called SPM, has also been successfully
applied to simulate properties of dispersions under several
non-equilibrium conditions, such as electrophoresis of charged spherical
particles under external electric fields \cite{kim1}. The method was
then modified to introduce thermal fluctuations into the dispersions
\cite{iwa1}, so that one can also simulate situations where the thermal
fluctuations and hydrodynamic interactions acting among dispersed
particles are both important. We have carried out systematic simulations
for dispersions composed of spherical particles with and without steady
shear flow, in order to analyze the diffusion process of dispersed
particles in detail \cite{iwa3} and to investigate the nonlinear
viscosity of the system \cite{iwa4}.

In the present paper, we report an important extension of the SPM for
analyzing the viscoelastic properties of particle dispersions immersed
in host fluids, by introducing a temporally oscillatory external force
into the system. To be consistent with the usual periodic boundary
conditions, shear flow with a zigzag profile was employed. The validity
of the method was examined by evaluating the storage $G'(\omega)$ and
loss $G''(\omega)$ moduli of a system comprising identical spherical
particles dispersed in an incompressible Newtonian host fluid at volume
fractions of $\Phi=0$, $0.41$, $0.46$, and $0.51$, for flow rates spanning a
range of frequencies, $0.005\leq \omega \leq 0.2$. We confirmed that the
method could successfully reproduce 1) purely viscous responses for
 $\Phi=0$ and 2) typical viscoelastic responses for $\Phi=0.41$ and $0.46$, 
in excellent agreement with experimental data obtained for
colloidal dispersions. 

To our knowledge, the present study is the first successful attempt to calculate the complex modulus of particle dispersions using DNS-type methods.
Further applications of our DNS method to more complex systems, such as
dispersions of non-spherical particles, particles in polymer matrices,
or dispersions of aggregating particles, are promising.

\end{document}